\documentclass[11pt]{article}
\usepackage{latexsym}  
\usepackage{amssymb}
\usepackage{graphicx}
\usepackage{amsmath}

\begin{document}

\begin{center}
{\Large\bf Maximal $CP$ Violation Hypothesis and 
a Lepton Mixing Matrix}

\vspace{4mm}
{\bf Yoshio Koide$^a$ and Hiroyuki Nishiura$^b$}

${}^a$ {\it Department of Physics, Osaka University, 
1-1 Machikaneyama, 
Toyonaka, Osaka 560-0043, Japan} \\
{\it E-mail address: koide@het.phys.sci.osaka-u.ac.jp}

${}^b$ {\it Faculty of Information Science and Technology, 
Osaka Institute of Technology, 
Hirakata, Osaka 573-0196, Japan}\\
{\it E-mail address: nishiura@is.oit.ac.jp}

\date{\today}
\end{center}

\vspace{3mm}
\begin{abstract}
Maximal $CP$ violation hypothesis is applied to 
a simple lepton mixing matrix form $U=V_{CKM}^\dagger U_{TB}$,
which has recently been speculated under an ansatz that 
$U$ becomes an exact tribimaximal mixing $U_{TB}$ in a limit 
of the quark mixing matrix $V_{CKM} \rightarrow {\bf 1}$.
The prediction $\tan^2 \theta_{12} =1/2$ in the case of 
the exact tribimaximal mixing $U=U_{TB}$ is considerably 
spoiled in the speculated mixing $U=V_{CKM}^\dagger U_{TB}$.
However, the application of the hypothesis 
to the lepton sector can again recover
the spoiled value to $\tan^2 \theta_{12} \simeq 1/2$ 
if the original Kobayashi-Maskawa phase convention for 
$V_{CKM}$ is adopted.
\end{abstract}

\vspace{3mm}

\noindent{\large\bf 1 \ Introduction}

Recently, an interesting form of the lepton mixing matrix
$U$ has been proposed \cite{KN-0809}:
$$
U=V^\dagger U_{TB} ,
\eqno(1.1)
$$
which was speculated under an ansatz that $U$ becomes an exact 
tribimaximal mixing \cite{tribi} $U_{TB}$ 
in a limit $V \rightarrow {\bf 1}$ ($V$ is the 
Cabibbo-Kobayashi-Maskawa (CKM) quark mixing matrix).
Here, $U_{TB}$ is given by  
$$
U_{TB} = P^\dagger(\gamma) U_{TB}^0 P(\sigma) ,
\eqno(1.2)
$$
where
$$
P(\gamma) ={\rm diag}(e^{i\gamma_1},e^{i\gamma_2},e^{i\gamma_3}),
\ \ \ 
P(\sigma) ={\rm diag}(e^{i\sigma_1},e^{i\sigma_2},e^{i\sigma_3}),
\eqno(1.3)
$$
$$
U_{TB}^0 = \left( 
\begin{array}{ccc}
\frac{2}{\sqrt6} & \frac{1}{\sqrt3} & 0 \\
-\frac{1}{\sqrt6} & \frac{1}{\sqrt3} & -\frac{1}{\sqrt2} \\
-\frac{1}{\sqrt6} & \frac{1}{\sqrt3} & \frac{1}{\sqrt2}
\end{array} \right) .
\eqno(1.4)
$$

A brief description of derivation of the relation (1.1) 
is as follows:  the up-quark and neutrino
mass matrices in the limit of $U_u \rightarrow {\bf 1}$
are given by $M_u^0=D_u$ and $M_\nu^0 =U_{TB} D_\nu U_{TB}^T$
($D_f ={\rm diag}(m_{f1}, m_{f2}, m_{f3})$), and those in
the observed world with a realistic small deviation 
$V \neq {\bf 1}$ from 
$V={\bf 1}$ become modified as 
$M_u^0 \rightarrow M_u=U_u D_u U_u^\dagger$ and 
$M_\nu^0 \rightarrow U_u M_\nu^0 U_u^\dagger$
(we use a mass matrix convention $U_f^\dagger M_f U_f =D_f$). 
Therefore, we obtain $U_\nu=U_u U_{TB}$ and
$U =U_e^\dagger U_\nu = U_e^\dagger U_d V^\dagger U_{TB}$,
which leads to the relation (1.1) by using an additional ansatz 
``$U_d \rightarrow {\bf 1}$ and $U_e \rightarrow {\bf 1}$
in the limit of $U_u \rightarrow {\bf 1}$" demanding approximately 
$U_e=U_d$. 
(For an explicit neutrino mass matrix model which gives the 
relation (1.1), 
see, for example, Refs.\cite{Rode05,Datta}.)
Note that we have assumed a hypothetical limit   
$V \rightarrow {\bf 1}$ which is realized by switching 
off terms giving $V \neq{\bf 1}$, irrespectively of
an energy scale.

The pure tribimaximal mixing $U=U_{TB}$ predicts 
$\tan^2\theta_{12}=1/2$, $\sin^2 2\theta_{23} =1$ and
$\sin^2\theta_{13}=0$ even if we consider a degree of freedom
due to the phase convention given by (1.2).
In contrast to the case $U=U_{TB}$, the predictions in the case 
$U=V^\dagger U_{TB}$
are spoiled by the presence of $P(\gamma)$.
Especially, the strict prediction $\tan^2\theta_{12}=1/2$ is 
considerably spoiled by the presence of a phase parameter 
$\beta\equiv \gamma_{ 2}-\gamma_{ 1}$: 
The predicted deviations of 
$\sin^2 2\theta_{23}$ and $\sin^2 \theta_{13}$ 
from those in the exact tribimaximal mixing $U=U_{TB}$
are small, i.e. $0.024 \leq \sin^2 \theta_{13} \leq  0.028$
and $0.94 \leq  \sin^2 2\theta_{23} \leq  0.95$  
depending on a phase parameter 
$\alpha\equiv \gamma_{ 3}-\gamma_{ 2}$,
while the prediction $\tan^2 \theta_{12}=1/2$ becomes vague, 
i.e. $0.24 \leq  \tan^2 \theta_{12} \leq  1.00$  
depending on the phase parameter  $\beta$
 (see Fig.3 in Ref.\cite{KN-0809}).
Here, the parameters $\alpha$ and $\beta$ are not 
observable parameters in the mixing matrix $U$,
but they are ``model-parameters".
However, since we fix the matrix $V$ in the ansatz (1.1)
by the observed CKM matrix parameters, the rotation 
angles and 
$CP$ violation phase parameter $\delta_\ell$ in the 
lepton mixing matrix
are completely determined by the parameters $\alpha$
and $\beta$ under the ansatz (1.1). 
(Note that the phase parameters $\sigma_i$,  which are 
the so-called Majorana phases, do not affect neutrino
oscillation phenomena.)
If we take $\beta \simeq \pi/2$,
we can again predict $\tan^2 \theta_{12} \simeq 1/2$.
This was pointed out by Plentinger and Rodejohann 
\cite{Rode05}, and also by the authors \cite{KN-0809}.
However, it is not clear whether 
the choice $\beta = \pi/2$ means really a case of the maximal $CP$ 
violation or not, because there are three $CP$ violating phases
in the present scenario, i.e.  
$\alpha$, $\beta$ and $\delta_q$ ($\delta_q$ is a $CP$ 
violating phase parameter in the CKM matrix $V(\delta_q)$). 

Since we apply the maximal $CP$ violation hypothesis
to the phenomenological ansatz (1.1), here, let us present 
a short review of the hypothesis.
Usually, the maximal $CP$ violation hypothesis is defined as
follows: the nature takes values of $CP$ violating phases 
so that a magnitude of the rephasing invariant quantity $J$ 
\cite{J} takes its maximal value.
Generally, the CKM matrix $V(\delta_q)$ is described by 4 
phase-convention-dependent parameters
(there are, in general, 9 phase conventions of the CKM 
matrix \cite{FX-9V}), i.e. three rotation 
parameters $(\theta_1, \theta_2,\theta_3)$  
and one $CP$ violating phase parameter $\delta_q$. 
We may choose the observable values $|V_{us}|$, $|V_{cb}|$, 
$|V_{ub}|$ and $|V_{td}|$ straightforwardly, instead of three 
rotation parameters and one phase parameter. 
In fact, as we demonstrate in the next section, 
we can fix the three rotation
parameters and one phase parameter from the observed 4 values 
of $|V_{ij}|$ when we adopt some phase convention
(but signs of the rotation parameters remain
as unsettled ones).
The rephasing invariant quantity $J$ is expressed by 
$J \propto \sin\delta_q$ in  any phase convention \cite{FX-9V}
of the CKM matrix, so that the maximal $CP$ 
violation means $\delta_q = \pi/2$. 
The requirement of this maximal
$CP$ violation, in general, put an over-constraint  on the
CKM parameters, because we already know the four independent
values of the CKM matrix $|V_{us}|$, $|V_{cb}|$, 
$|V_{ub}|$ and $|V_{td}|$.
As we demonstrate in the next section, we find that 
only the original Kobayashi-Maskawa (KM) phase convention
\cite{CKM_KM} can satisfies the maximal $CP$ violation 
hypothesis \cite{YK-maxCP}.
We know that the physics in the CKM mixing are invariant
under the rephasing.
On the other hand, we know that the phase conventions of 
the CKM matrix are deeply related to explicit mass 
matrix forms in the models.
This suggests that the hypothesis is not  
for parameters in the CKM matrix, but for those in a mass 
matrix model. 
It should be noted that the maximal $CP$ violation 
hypothesis is not one based on a theoretical ground but
a phenomenological one.

In the present paper, we extend the maximal $CP$ violation
hypothesis to the following hypothesis:
When the rephasing invariant quantity $J$ is a function of $CP$ 
violating phases $\delta_1$, $\delta_2$, $\cdots$, i.e.
$J=J(\delta_1, \delta_2, \cdots)$, the maximal $CP$ violation
hypothesis requires 
$$
\frac{\partial J}{\partial \delta_1} = 
\frac{\partial J}{\partial \delta_2} = \cdots = 0 ,
\eqno(1.5)
$$
under the condition that rotation parameters are fixed.
Here, $\delta_1$, $\delta_2$, $\cdots$ are $CP$ violating
phase parameters in a mass matrix model.
Note that the mixing matrix $V$ ($U$) can always be 
expressed by three rotation parameters and one phase parameter 
$\delta_q$ ($\delta_\ell$), and they can become observable
parameters when we adopt some phase convention.
In contrast to these four parameters in the mixing matrix, 
the phases $\delta_i$ are not observable even when 
we adopt a phase convention.
The $CP$ violating parameter $\delta_q$  ($\delta_\ell$)
is given by a function of $\delta_i$ and 
other mass matrix parameters.
By abbreviating $\delta_q$ ($\delta_\ell$) to 
$\delta$ we have 
$$
\frac{\partial J}{\partial \delta}= 
\frac{\partial J}{\partial \delta_1} 
\frac{\partial \delta_1}{\partial \delta}+
\frac{\partial J}{\partial \delta_2} 
\frac{\partial \delta_2}{\partial \delta}+ \cdots .
\eqno(1.6)
$$
Therefore,@it turns out that the requirement (1.5) 
is considerably stronger than
the constraint $\partial J/\partial \delta =0$.

First, let us demonstrate that
even when $J$ involves only one $CP$ violating
phase $\delta$, results based on the above definition 
of the maximal $CP$ violation hypothesis depend on phase
conventions of the flavor mixing matrix \cite{YK-maxCP}.
For example, in the standard expression \cite{CKM_SD} 
$V_{SD}(\delta_{SD})$ 
and original Kobayashi-Maskawa (KM) expression
\cite{CKM_KM} 
$V_{KM}(\delta_{KM})$ of $V$, the rephasing invariant quantity 
$J$ is given by
$$
J_{SD} =c_{13}^2 s_{13} c_{12}s_{12} c_{23}s_{23}\sin\delta_{SD} ,
\eqno(1.7)
$$
and
$$
J_{KM}=c_1 s_1^2 c_2 s_2 c_3 s_3 \sin\delta_{KM} ,
\eqno(1.8)
$$
respectively.
Here, $V_{SD}$ and $V_{KM}$ are explicitly given by
$$
V_{SD} = R_1 (\theta_{23}) P_3(\delta_{SD}) R_2 (\theta_{13}) 
P_3^{\dagger} (\delta_{SD}) R_3 (\theta_{12})
$$
$$
= \left(
\begin{array}{ccc}
c_{13} c_{12} & c_{13} s_{12} & s_{13} e^{-i\delta_{SD}} \\
-c_{23} s_{12} -s_{23} c_{12} s_{13} e^{i\delta_{SD}} &
c_{23} c_{12} -s_{23} s_{12} s_{13} e^{i\delta_{SD}} &
s_{23} c_{13} \\
s_{23} s_{12} -c_{23} c_{12} s_{13} e^{i\delta_{SD}} &
-s_{23} c_{12} -c_{23} s_{12} s_{13} e^{i\delta_{SD}} &
c_{23} c_{13} 
\end{array} \right) ,
\eqno(1.9)
$$
$$
V_{KM} = R_1^{T} (\theta_2) P_3 (\delta_{KM} + \pi) R_3 (\theta_1)
R_1 (\theta_3)
$$
$$
= \left(	
\begin{array}{ccc}
c_{1}  & -s_{1} c_{3} & -s_{1} s_3  \\
s_1 c_2 & c_1 c_2 c_3 -s_2 s_3 e^{i\delta_{KM}} &
c_1 c_2 s_3 + s_2 c_3 e^{i \delta_{KM}} \\
s_1 s_2 & c_1 s_2 c_3 +c_2 s_3 e^{i\delta_{KM}} &
c_1 s_2 s_3 -  c_2 c_3 e^{i \delta_{KM}}
\end{array} \right) ,
\eqno(1.10)
$$
respectively, where
$$
R_1 (\theta) = \left(
\begin{array}{ccc}
1 & 0 & 0 \\
0 & c & s \\
0 & -s & c 
\end{array} \right) , \ \ \ \ 
R_2 (\theta) = \left(
\begin{array}{ccc}
c & 0 & s \\
0 & 1 & 0 \\
-s & 0 & c 
\end{array} \right) , \ \ \ \ 
R_3 (\theta) = \left(
\begin{array}{ccc}
c & s & 0 \\
-s & c & 0 \\
0 & 0 & 1 
\end{array} \right) ,
\eqno(1.11)
$$
$$
P_3 (\delta) = {\rm diag} (1, \ 1, \ e^{i \delta}), \ \ \ 
\eqno(1.12)
$$
$s = \sin \theta$ and  $c = \cos \theta$.
It is well known \cite{YK-maxCP} that the standard expression 
$V_{SD}(\delta_{SD})$ with $\delta_{SD}=\pm \pi/2$
cannot describe the observed CKM matrix parameters,
while $V_{KM}(\delta_{KM})$ with $\delta_{KM}=\pm \pi/2$
can well describe the observed those.
(In the standard phase convention $V_{SD}(\delta_{SD})$, 
a case with $\delta_{SD} \simeq 70^\circ$ is in favor of the 
observed data.)
Thus the requirement of the maximal $CP$ violation in the quark 
sector can give a reasonable value for the CKM phase parameter 
only when the original KM matrix phase convention is adopted. 
From such a phenomenological point of view,
we adopt this convention not only for the quark sector 
but also for the lepton sector [i.e. $V$ in the lepton mixing
matrix $U$ given by Eq.(1.1)] in order to ensure consistency.

In this paper, we assume the maximal $CP$ violation hypothesis
for both the quark and lepton sectors.
In this scenario, since the matrix $V(\delta_q)$ in
Eq.(1.1) is already fixed by the observed data in the quark 
sector, the rephasing invariant quantity $J$ is only a 
function of $\alpha$ and $\beta$.
In Sec.2, we re-investigate the CKM mixing parameters
from the data in the quark sector, and fix mixing parameters
in the phase convention $V=V_{KM}$ at $\mu \simeq m_Z$ 
under the maximal $CP$ violation hypothesis.
Here we use an energy scale $\mu=m_Z$ at which the maximal 
$CP$ violation hypothesis in the quark sector seems 
to work out.
In Sec.3, we will apply the maximal $CP$ violation hypothesis
to the lepton mixing $U=V^\dagger U_{TB}$ with $V=V_{KM}$.
We find that the maximal value of $|J(\alpha, \beta)|$
takes place at $\beta \simeq \pm \pi/2$ and $\alpha \simeq 0$
(or $\alpha \simeq \pi$), so that we can again obtain 
$\tan^2 \theta_{12} \simeq 1/2$.
(Note that the definition of the parameter $\alpha$ and 
$\beta$ in the present paper are different from those
in the previous paper \cite{KN-0809}, because the CKM
matrix $V$ in $U=V^\dagger U_{TB}$ was $V_{SD}$ in the 
previous paper, while the present one is $V_{KM}$.)
Finally, Sec.4 is devoted to the summary and concluding 
remarks.


\vspace{3mm}

\noindent{\large\bf 2 \ Maximal $CP$ violation hypothesis 
in the quark sector}

First, we estimate the CKM matrix parameters in the 
original KM matrix $V_{KM}(\delta_{KM})$
without assuming the maximal $CP$ violation.
Using input values \cite{PDG08} 
$|V_{us}|=0.2255 \pm 0.0019$,  $|V_{ub}|=0.00393 \pm 0.00036$ 
and $|V_{td}|=0.0081\pm 0.0006$, 
we obtain the rotation parameters
$$
|s_1| = 0.2255 \pm 0.0019, \ \ |s_2| =0.0359^{+0.0030}_{-0.0029},
 \ \ |s_3|=0.0174^{+0.0018}_{-0.0017} .
\eqno(2.1)
$$
By fitting the value of $\delta_{KM}$ to the observed value 
$|V_{cb}|=0.0412 \pm 0.0011$, we obtain  
$\delta_{KM} =(84^{+16}_{-22})^\circ$. 
The present observed values do not give an exact value 
$\delta_{KM}=\pi/2$, but it is not ruled out.

Inversely, if we assume the maximal $CP$ violation, i.e.
$\delta_{KM}=\pm \pi/2$, we can
fix the parameters $s_1$, $s_2$ and $s_3$ from the observed
values of $|V_{us}|$, $|V_{cb}|$ and $|V_{ub}|$, and can
predict the value of $|V_{td}|$. 
(Although the value of $s_2$ is readily fixed from
the relation $V_{td}= s_1 s_2$ in the original KM matrix, 
we use the value $|V_{cb}|$ as an input value, because
the accuracy of $|V_{td}|$ is not so precise compared with
that of $|V_{cb}|$.)
For convenience, we define $V_{us} > 0$, so that we take 
$s_1= -\sqrt{|V_{us}|^2+|V_{ub}|^2} <0$ and
$s_3=V_{ub}/\sqrt{|V_{us}|^2+|V_{ub}|^2}$.
We also define that all $c_i$ ($i=1,2,3$) are positive,
i.e. $c_i =\sqrt{1-s_i^2} >0$.
For input values \cite{PDG08}, $V_{us}=0.2255\pm 0.0019$,
$V_{ub}=-s_1 s_3 = \pm (0.00393\pm 0.00036)$, 
$|V_{cb}| =0.0412\pm 0.0011$, we obtain reasonable 
CKM parameter values only for cases with $s_3 s_\delta/s_2 >0$
($s_\delta =\sin\delta_{KM}=\pm 1$): 
$$
s_1 = -(0.2255 \pm 0.0019), \ \
|s_2| = 0.0376^{+0.0019}_{-0.0021} , \ \  
|s_3|=0.0174^{+0.0018}_{-0.0017} ,
\eqno(2.2)
$$
$$
|V_{td}|=0.0085 \pm 0.0005 , 
\eqno(2.3)
$$
$$
\phi_1=(24.4^{-3.5}_{+3.2})^\circ ,\ \
\phi_2=(89.963 \mp 0.004)^\circ ,\ \ 
\phi_3=(65.7^{+3.1}_{-3.5})^\circ ,
\eqno(2.4)
$$
where 
the angles $\phi_i$ of the unitary triangle have been
defined by
$$
\phi_1={\rm arg}\left(-\frac{V_{21}V_{23}^*}{V_{31}V_{33}^*}
\right) , \ \ 
\phi_2={\rm arg}\left(-\frac{V_{31}V_{33}^*}{V_{11}V_{13}^*}
\right) , \ \ 
\phi_3={\rm arg}\left(-\frac{V_{11}V_{13}^*}{V_{21}V_{23}^*}
\right) . 
\eqno(2.5)
$$ 
Those predicted values are in agreement with the
observed CKM matrix data \cite{PDG08}.  

Next we consider the case in which we adopt $V=V_{SD}(\delta_{SD})$  
instead of using $V_{KM}$.  In this standard phase convention, 
by using the global fit values, $|V_{us}| =0.2257\pm 0.0010$, 
$|V_{cb}|=0.0415^{+0.0010}_{-0.0011}$, 
$|V_{ub}|=0.00359 \pm 0.00016$ and
$|V_{td}|=0.00874^{+0.00026}_{-0.00037}$,
reported by Particle Data Group \cite{PDG08}, we obtain
$$
\delta_{SD}=(68.9^{+\ 9.1}_{-10.7})^\circ . \ \ 
\eqno(2.6)
$$
Thus, for the standard phase convention $V_{SD}$, we cannot
demand the maximal $CP$ violation hypothesis consistently, 
because the value $\delta_{SD}=(68.9^{+\ 9.1}_{-10.7})^\circ$
is far from the value $\delta_{SD}=\pi/2$ in the maximal 
$CP$ violation hypothesis.

\vspace{3mm}

\noindent{\large\bf 3 \ Maximal $CP$ violation hypothesis 
in the lepton sector}

We assume that the lepton mixing matrix $U$ is given by Eq.(1.1).
Although the observable parameters in the matrix $U$ are 
three rotation parameters and one phase parameter, we 
practically have two parameters 
$\alpha \equiv \gamma_{ 3}-\gamma_{ 2}$ and 
$\beta \equiv \gamma_{ 2} -\gamma_{ 1}$  
as adjustable parameters,  
because we fix the values of the CKM 
matrix $V$ by the observed one $V=V_{KM}$.
We apply the ansatz (1.5) to the lepton 
mixing matrix $U$ with the free parameters 
$\alpha$ and $\beta$. 
The parameters $\alpha$ and $\beta$ correspond 
to $\delta_1$ and $\delta_2$ in Eq.(1.5).
Of course, the observable parameter in $CP$ violation
is only $\delta_\ell$ in the present model (1.1), 
although it is not explicitly given in the present
paper.

Now, we calculate the rephasing invariant quantity $J$
in the lepton sector using the relation
$$
J = {\rm Im}(U_{23} U_{12} U_{22}^* U_{13}^*),
\eqno(3.1)
$$
where
$$
\begin{array}{l}
U_{12} =\frac{1}{\sqrt3} \left( V_{11}^* e^{-i \gamma_{ 1}}
+V_{22}^* e^{-i \gamma_{ 2}}+V_{32}^* e^{-i \gamma_{ 3}}
\right) e^{i\sigma_2} , \\
U_{22}^* =\frac{1}{\sqrt3} \left( V_{12} e^{i \gamma_{ 1}}
+V_{21} e^{i \gamma_{ 2}}+V_{31} e^{i \gamma_{ 3}}
\right) e^{-i\sigma_2} , \\
U_{23} =\frac{1}{\sqrt2} \left( 
-V_{22}^* e^{-i \gamma_{ 2}}+V_{32}^* e^{-i \gamma_{ 3}}
\right) e^{i\sigma_3} , \\ 
U_{13}^* =\frac{1}{\sqrt2} \left( 
-V_{21} e^{i \gamma_{ 2}}+V_{31} e^{i \gamma_{ 3}}
\right) e^{-i\sigma_3} . 
\end{array}
\eqno(3.2)
$$
Here, the lepton mixing matrix $U$ is given by the form
(1.1), i.e. $U=V^\dagger U_{TB}$.
Note that the CKM mixing matrix $V$ should be estimated at energy 
scale $\mu \simeq m_Z$@by using a specific phase convention.
Since we assume the maximal $CP$ violation hypothesis for
the quark sectors, too, we adopt the CKM matrix parameters
$\theta_1$, $\theta_2$, $\theta_3$ and $\delta_{KM}=\pm \pi/2$ 
in the original KM phase convention as we discussed in the
previous section.
Since the numerical results for the mixing $U$ are dependent
on the phase convention of $V$, the predicted values of the
neutrino mixing parameters in the present paper are different
from those in the previous paper \cite{KN-0809}, where 
the phase convention $V=V_{SD}$ was adopted.
For reference, we illustrate the numerical results of 
the neutrino mixing parameters $\sin^2 \theta_{13}$, 
$\sin^2 2\theta_{23}$ and $\tan^2 \theta_{12}$ in Figs.1-3,
correspondingly to Figs.1-3 in the previous paper.
Although the numerical results are almost similar to the
previous ones, a value $(\alpha, \beta)$ which gives a maximal 
$|J|$ is changed from the previous one.


{\scalebox{1.0}{\includegraphics{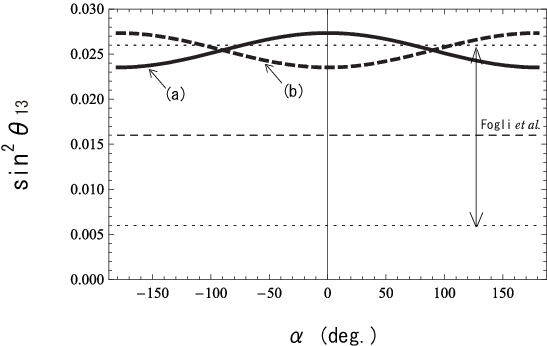}} }

\begin{quotation}\small
{\bf Fig.~1} \  Behavior of $\sin^2 \theta_{13}$
versus $\alpha$.
Curves are drawn for inputs (a)  
$(s_\delta, s_1, s_2, s_3) = (+1, -0.2255, -0.0376, -0.0174)$
and (b) $(s_\delta, s_1, s_2, s_3) =(+1, -0.2255, +0.0376, +0.0174)$,
where $s_\delta =\sin\delta_{KM}$.
\end{quotation}


{\scalebox{1.0}{\includegraphics{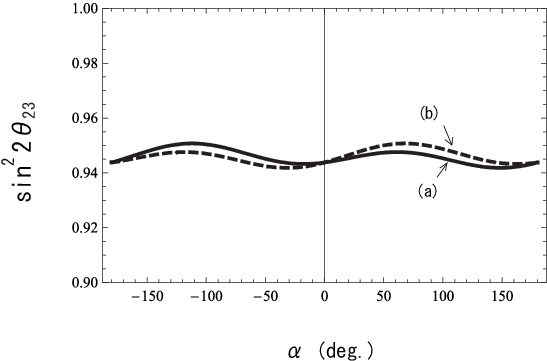}} }

\begin{quotation}\small
{\bf Fig.~2}  Behavior of $\sin^2 2\theta_{23}$
versus $\alpha$. 
Curves are drawn for inputs (a) 
$(s_\delta, s_1, s_2, s_3) = (+1, -0.2255, -0.0376, -0.0174)$
and (b) $(s_\delta, s_1, s_2, s_3) =(+1, -0.2255, +0.0376, +0.0174)$,
where $s_\delta =\sin\delta_{KM}$.
\end{quotation}


{\scalebox{1.0}{\includegraphics{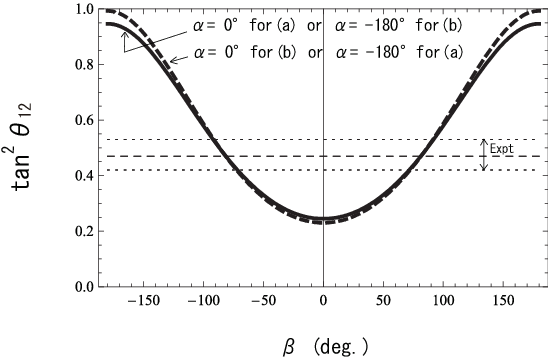}} }

\begin{quotation}\small
{\bf Fig.~3}  Behavior of $\tan^2\theta_{12}$
versus $\beta$ for typical values of $\alpha$. 
Curves are drawn by taking $\alpha=0^{\circ}$ and $-180^\circ$ 
for inputs (a) 
$(s_\delta, s_1, s_2, s_3)=(+1, -0.2255, -0.0376, -0.0174)$
and (b)
$(s_\delta, s_1, s_2, s_3)=(+1, -0.2255, +0.0376, +0.0174)$,
where $s_\delta =\sin\delta_{KM}$.
 
\end{quotation}

From Eq.(2.1), we obtain
$$
J \simeq \frac{1}{6} s_1 ( s_\beta + s_2  s_\alpha c_\beta
-s_2  c_\alpha s_\beta +2 s_3 c_\alpha c_\beta s_\delta)  ,
\eqno(3.3)
$$
where $\alpha=\gamma_{ 3}-\gamma_{ 2}$, 
$\beta=\gamma_{ 2}-\gamma_{ 1}$, 
$s_\delta=\sin\delta_{KM} =\pm 1$,  
and $c_\alpha=\cos\alpha$ and so on, and we have used
the observed fact $1\gg |s_1|\simeq |V_{us}| \gg |s_2| 
\simeq |V_{td}|/|V_{us}| \sim |s_3|\simeq |V_{ub}|/|V_{us}|$.
The value $J$ is approximately given by 
$J\simeq (1/6)\sin\theta_1 \sin\beta$, so that
the maximal $CP$ violation hypothesis demands
$\beta \simeq \pm \pi/2$ (however, the small deviation
from $\beta=\pm \pi/2$ is crucial).
More precisely speaking, from $\partial J/\partial \beta=0$,
we obtain
$$
\cot\beta \simeq  2 s_3 c_\alpha s_\delta +s_2 s_\alpha .
\eqno(3.4)
$$
Similarly, we obtain
$$
\tan\alpha \simeq \frac{s_2 c_\beta}{2 s_3 c_\beta s_\delta 
-s_2 s_\beta} ,
\eqno(3.5)
$$
from $\partial J/\partial \alpha=0$
(but with a rough approximation).
Since $\beta\simeq \pm \pi/2$ from Eq.(3.4), we obtain 
$\alpha \simeq 0$ or $\pi$ from Eq.(3.5).
We emphasize that the maximal $CP$ violation hypothesis
can determine values of the phase parameters $\alpha$ and 
$\beta$ simultaneously. 
The numerical results obtained with use of no
approximation are given in Table 1.
As an example of the behavior of $|J(\alpha,\beta)|$,
$J$ versus $\alpha$ in a typical case $(s_\delta,s_2,s_3)=(+,-,-)$
in Table 1 is illustrated in Fig.~4.
As seen in Table 1, the value of $\alpha$ takes $0$ or $\pi$
according as $s_2 <0$ or $s_2> 0$, i.e. $V_{td} <0$ or 
$V_{td} > 0$.
For comparison, we show the results for the case of $V=V_{SD}$ 
in Table 2.
In this case, we obtain $\alpha \simeq 25^\circ$,
although we can still obtain $\beta \simeq \pi/2$.

\begin{table}

\begin{tabular}{|c|c|c|c|c|c|c|} \hline
$s_\delta$ & $s_2$ & $s_3$ & $(\pm J)_{max}$ & $\alpha$ & $\beta$ 
& Case \\ \hline
 $+$ & $+$ & $+$ & $0.03772\pm 0.00037$ & $-(175.6^{-1.1}_{+0.4})^\circ$ 
& $-(87.93 \mp 0.21)^\circ$ & A$_1$ \\
& & & $-(0.03800^{-0.00037}_{+0.00034})$ & $+(179.3^{+0.6}_{-1.3})^\circ$ 
& $+(91.88^{+0.20}_{-0.22})^\circ$ & A$_2$ \\ \hline
 $+$ & $-$ & $-$ & $0.03772\pm 0.00037$ & $+(3.64^{+0.46}_{-0.44})^\circ $ 
& $-(87.96 \mp 0.21)^\circ $ & B$_1$ \\
&  & & $-(0.03801^{-0.00035}_{+0.00034})$ & $+(2.86^{+0.29}_{-0.30})^\circ$ 
& $+(92.01^{+0.21}_{-0.20})^\circ$ & B$_2$ \\ \hline
 $-$ & $+$ & $-$ & $0.03772\pm 0.00037$ & $-(175.6^{-1.1}_{+0.4})^\circ$ 
& $-(87.93 \mp 0.21)^\circ$ & A$_1$ \\
& & & $-(0.03800^{-0.00037}_{+0.00034})$ & $+(179.3^{+0.6}_{-1.3})^\circ$ 
& $+(91.88^{+0.20}_{-0.22})^\circ$ & A$_2$ \\ \hline
 $-$ & $-$ & $+$ & $0.03772\pm 0.00037$ & $+(3.64^{+0.46}_{-0.44})^\circ $ 
& $-(87.96 \mp 0.21)^\circ $ & B$_1$ \\
&  & & $-(0.03801^{-0.00035}_{+0.00034})$ & $+(2.86^{+0.29}_{-0.30})^\circ$ 
& $+(92.01^{+0.21}_{-0.20})^\circ$ & B$_2$ \\ \hline
\end{tabular}

\caption{\small Possible solutions of $CP$ 
violating phase factors 
$\alpha=\gamma_{ 3}-\gamma_{ 2}$ and 
$\beta=\gamma_{ 2}-\gamma_{ 1}$
under the maximal $CP$ violation hypothesis. 
We obtain four sets of ($\alpha$, $\beta$), which are denoted 
by A$_1$, A$_2$, B$_1$, and B$_2$.
}
\end{table}


{\scalebox{1.0}{\includegraphics{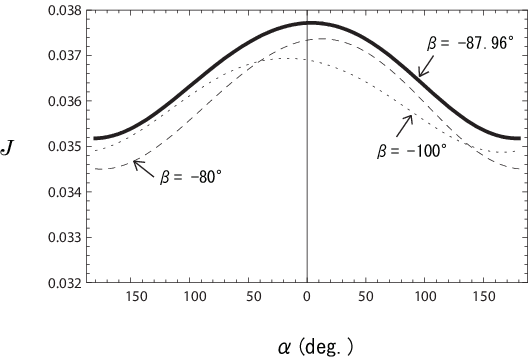}} }

\begin{quotation}\small
{\bf Fig.~4}  An example of the behavior of $J(\alpha,\beta)$
versus $\alpha$ for typical values of $\beta$. 
The case corresponds to the case 
with $(s_\delta, s_2, s_3)=(+,-,-)$ in Table 1. 
\end{quotation}

\begin{table}

\begin{tabular}{|c|c|c|} \hline
 $(\pm J)_{max}$ & $\alpha$ & $\beta$ \\ \hline
$+(0.0378 \pm 0.0002)$ & $+(25.4^{+2.4}_{-3.6})^\circ$ & 
$-(88.2^{+0.1}_{-0.2})^\circ$ \\
$-(0.0381 \pm 0.0002)$ & $+(24.5^{+1.8}_{-1.7})^\circ$ & 
$+(91.8 \pm 0.1)^\circ$ \\
\hline
\end{tabular}

\caption{\small Possible values of $CP$ violating phase factors 
$\alpha=\gamma_{ 3}-\gamma_{ 2}$ and 
$\beta=\gamma_{ 2}-\gamma_{ 1}$ for the case
$V=V_{SD}(\delta_{SD})$ with $\delta_{SD} =(68.9^{+\ 9.1}_{-10.7})^\circ$.
}
\end{table}


\vspace{3mm}
\vspace{3mm}

\noindent{\large\bf 4 \ Summary}

In conclusion, we have applied an extended ``maximal CP 
violation hypothesis" (1.5) to a simple lepton mixing 
matrix form $U=V^\dagger U_{TB}$, which has recently 
been speculated under an ansatz that 
$U$ becomes an exact tribimaximal mixing $U_{TB}$ in a limit 
of the quark mixing matrix $V \rightarrow {\bf 1}$.
The mixing matrix $U_{TB}$ includes two phase parameters
$\alpha=\gamma_{ 3} -\gamma_{ 2}$ and
$\beta=\gamma_{ 2} -\gamma_{ 1}$ due to the
phase convention of the tribimaximal mixing.
Therefore, the rephasing invariant quantity $J$ in the
lepton sector is a function of phase parameters $\alpha$,
$\beta$ and $\delta_q$ ($\delta_q$ is a $CP$ violating
phase parameter in the quark mixing matrix $V(\delta_q)$).
We have demanded the maximal $CP$ violation 
hypothesis for the quark sector too. 
Thus, we have taken the original KM phase convention 
$V_{KM}(\delta_{KM})$ with $\delta_{KM}= \pm \pi/2$ 
as the CKM matrix $V$ in Eq.(1.1),
because the standard phase convention $V=V_{SD}(\delta_{SD})$
with $\delta_{SD}=\pm \pi/2$ cannot reproduce the observed 
CKM parameters consistently under the hypothesis (1.5).
Then, the quantity $J$ in the lepton sector is a 
function of only $\alpha$ and $\beta$.
We have regarded the parameters $\alpha$ and $\beta$ as 
the independent $CP$ violation parameters in applying 
the maximal $CP$ violation hypothesis (1.5) to the 
lepton mixing matrix (1.1), although the observable 
$CP$ violation parameter is still a parameter
$\delta_\ell$ which is given by a function of 
$\alpha$ and $\beta$. 
(For example, we can choose 
$\sin^2 2\theta_{23}$, $\tan^2 \theta_{12}$, 
$|U_{13}|$ and $\delta_\ell$ as the four observable 
quantities in the lepton mixing matrix $U$ 
except for Majorana phase parameters.)

We have found that 
only for the case $V=V_{KM}$, the maximal 
$CP$ violation hypothesis leads to interesting results,
$\delta_{KM} =\pm \pi/2$ in the quark sector, and
$\beta \simeq  \pm \pi/2$ and $\alpha \simeq 0$ (Cases A$_1$ 
and A$_2$)
[or $\alpha \simeq \pi$ (Cases B$_1$ and B$_2$)] in the 
lepton sector.
The result $\beta\simeq \pm \pi/2$ predicts \cite{KN-0809}
$\tan^2 \theta_{12} \simeq 1/2$ which is in good
agreement with the observed value $\tan^2 \theta_{12}
=0.47^{+0.05}_{-0.04}$ \cite{SNO08}.
The result $\alpha \simeq 0$ (or $\alpha \simeq \pi$) 
means that the neutrino mass matrix $M_\nu^0=
U_{TB} D_\nu U_{TB}^T$ in the limit of $V \rightarrow 
{\bf 1}$ is nearly $2\leftrightarrow 3$ symmetric
(or antisymmetric). 
The predicted neutrino oscillation 
parameters are listed in Table 3 for the possible cases
defined in Table 1.
The predicted values are consistent with the 
observed values $\sin^2 2 \theta_{23} =1.00_{-0.13}$ 
\cite{MINOS}, $\tan^2 \theta_{12}=0.47^{+0.05}_{-0.04}$ 
\cite{SNO08} and $\sin^2 \theta_{13}=0.016\pm 0.010$
($1\sigma$) \cite{Fogli08}, although the predicted value
$\sin^2 \theta_{13}=0.0273$ is somewhat critical compared
with the value \cite{Fogli08} $\sin^2 \theta_{13}=0.016+0.010$ 
reported by Fogli {\it et al}.

\begin{table}

\begin{tabular}{|c|c|c|c|} \hline
Case & $\sin^2 2\theta_{23}$ & $\sin^2\theta_{13}$ & 
$\tan^2\theta_{12}$ \\ \hline
A$_1$ & $0.944\pm 0.001$ & 
$0.0273\pm 0.0005$ & $0.507 \pm 0.001$  \\
A$_2$  &  $0.944\pm 0.001$   &
$0.0273\pm 0.0005$    &  $0.530 \pm 0.001$   \\ \hline
B$_1$  &  $0.944\pm 0.001$   &
$0.0273\pm 0.0006$  &  $0.5083 \pm 0.0003$  \\
B$_2$  &  $0.944\pm 0.001$   & 
$0.0273\pm 0.0006$ &  $0.529 \pm 0.001$  \\ \hline
\end{tabular}

\caption{\small Predicted values of neutrino oscillation
parameters in the cases defined in Table 1. 
}
\end{table}

It is worthwhile noticing that the neutrino mixing
matrix $U=V^\dagger U_{TB}$ with the realistic $V\neq {\bf 1}$
spoils the prediction
$\tan^2\theta_{12}=1/2$ in the pure tribimaximal mixing
$U=U_{TB}$ as $0.24 \leq \tan^2\theta_{12} \leq 1.00$,
while the maximal $CP$ violation hypothesis fixes 
the phase parameter $\beta$ as $\beta\simeq \pm \pi/2$,
so that the hypothesis recovers the spoiled value of 
$\tan^2\theta_{12}$ to
$\tan^2\theta_{12} \simeq 1/2$.
The parameter $\beta$  is fixed almost independently
of the phase convention of the quark mixing matrix $V$,
while the parameter $\alpha$ is fixed dependently 
on the phase convention of $V$: 
If we take $V=V_{KM}(\delta_{KM})$
with $\delta_{KM}=\pm \pi/2$ under the maximal $CP$
violation hypothesis, we obtain the result 
$\alpha\simeq 0$ or $\pi$. On the other hand, if we take
$V=V_{SD}(\delta_{SD})$
with $\delta_{SD}=68.9^\circ$ (without the maximal $CP$
violation hypothesis in the quark sector),  we obtain 
$\alpha \simeq 25^\circ$,
which does not seem to be a suggestive value.
Thus, the maximal $CP$ violation hypothesis can lead to 
phenomenologically interesting results not only in the quark 
sector, but also in the lepton sector. 
However, the reason why the hypothesis is so effective
only when we take $V=V_{KM}$ has still not been understood. 
Also, theoretical ground for the maximal $CP$ violation 
hypothesis has still been unclear. 
We hope that, by investigating these problems, 
one will find a promising clue to a unified mass matrix model.

\vspace{3mm}

\centerline{\large\bf Acknowledgment}

One of the authors (YK) is supported by the Grant-in-Aid for
Scientific Research, JSPS (No.18540284 and No.21540266).

\vspace{4mm}


\begin{thebibliography}{99}
%
\bibitem{KN-0809} Y.~Koide and H.~Nishiura, Phys.~Lett. 
{\bf B669},  24 (2008).
%
\bibitem{tribi} 
P.~F.~Harrison, D.~H.~Perkins and W.~G.~Scott,
 Phys.~Lett. {\bf B458}, 79 (1999);
 Phys.~Lett. {\bf B530} (2002) 167;
Z.-z.~Xing, Phys.~Lett. {\bf B533}, 85 (2002);
P.~F.~Harrison and W.~G.~Scott,  Phys.~Lett. {\bf B535}, 163 (2002);
Phys.~Lett. {\bf B557}, 76 (2003);
E.~Ma, Phys.~Rev.~Lett. {\bf 90}, 221802 (2003);
C.~I.~Low and R.~R.~Volkas, Phys.~Rev. {\bf D68}, 033007 (2003).
%
\bibitem{Rode05}
F.~Plentinger and W.~Rodejohann, Phys.~Lett. {\bf B625}, 264 (2005).
%
\bibitem{Datta}
A.~Datta, arXiv:0807.0420 [hep-ph]; Phys.~Rev. {\bf D78}, 095004 (2008).
%
\bibitem{J}
C.~Jarlskog, Phys.~Rev.~Lett. {\bf 55}, 1839 (1985);
O.~W.~Greenberg,  Phys.~Rev. {\bf D32}, 1841 (1985);
I.~Dunietz, O.~W.~Greenberg and D.-d.~Wu,  Phys.~Rev.~Lett. 
{\bf 55}, 2935 (1985);
C.~Hamzaoui and A.~Barroso,  Phys.~Rev. {\bf D33}, 860 (1986).
%
\bibitem{FX-9V}
H.~Fritzsch and Z.~-z.~Xing, Phys.~Rev. {\bf D57}, 594 (1998).
Also, see Y.~Koide, Phys.~Rev. {\bf D73}, 073002 (2006).
%
\bibitem{CKM_KM}
M.~Kobayashi and T.~Maskawa, Prog.~Theor.~Phys. {\bf 49}, 652
(1973).
%
\bibitem{YK-maxCP} Y.~Koide, Phys.~Lett. {\bf B607}, 123 (2005).
%
\bibitem{CKM_SD}
L.~-L.~Chau and W.~-Y.~Keung, Phys.~Rev.~Lett. {\bf 53}, 1802 (1984);
H.~Fritzsch, Phys.~Rev. {\bf D32}, 3058 (1985).
%
\bibitem{PDG08} Particle Data Group, Phys.~Lett. {\bf B667}, 1 (2008).
%
\bibitem{SNO08} B.~Aharmim, {\it et al.}, SNO collaboration,
Phys.~Rev.~Lett. {\bf 101}, 111301 (2008).
Also, see S.~Abe, {\it et al.}, KamLAND collaboration,
Phys.~Rev.~Lett. {\bf 100}, 221803 (2008).
%
\bibitem{MINOS} D.~G.~Michael {\it et al.}, MINOS collaboration,
Phys.~Rev.~Lett. {\bf 97}, 191801 (2006);
J.~Hosaka, {\it et al.}, Super-Kamiokande collaboration, Phys.~Rev. 
{\bf D74} 032002 (2006).
%
\bibitem{Fogli08}
G.~Fogli, E.~Lisi, A.~Marrone, A.~Palazzo and A.~M.~Rotunno,
Phys.~Rev.~Lett. {\bf 101}, 141801 (2008).
%
\end{thebibliography}
\end{document}